\documentclass{osa-article}

\journal{osajournal}

\articletype{Research Article}

\begin{document}

\title{Lensing properties of rotational gas flow}

\author{D. Kaganovich$^*$, L. A. Johnson, D. F. Gordon, A. A. Mamonau$^1$, and B. Hafizi}

\address{
Plasma Physics Division, Naval Research Laboratory, Washington, D.C. 20375, 
\authormark{1}Also with Research Support Instruments, Lanham, MD 20706} 

\email{\authormark{*}dmitri.kaganovich@nrl.navy.mil} 

\begin{abstract*}
A negative lens comprised of a gas in steady axisymmetric flow is demonstrated experimentally and analyzed. The lens has potential applications in high-intensity laser optics and presents the possibility of adjusting the focusing properties of the lens on a sub-millisecond time-scale. It can be operated in environments where conventional optical elements are vulnerable. 
\end{abstract*}

\section{Introduction}
In applications requiring down-collimation of high-intensity laser beams the most vulnerable optical component is the output lens of the collimator, where the laser intensity is the highest. The damage threshold of this lens is typically the limiting factor \cite{Said_95, Couairon_05} for the beam size in laser propagation experiments where a small-diameter, collimated laser beam is required. This limitation can be mitigated by a negative (defocusing) lens made out of gas. 

Positive (focusing) gas lenses were demonstrated in long, heated, spinning tubes \cite{Notcutt_88}. These lenses were used for imaging distant objects \cite{Michaelis_91} and proposed for applications in astronomy. Focusing of laser beams for metal cutting applications was also successfully demonstrated in heated tube lenses \cite{Michaelis_86}. In spinning tube gas lenses the diameter of the lens is typically smaller than its length, limiting the field of view and speed of the lens \cite{Mafusire_08}.

In this paper we demonstrate a neutral gas negative lensing structure based on the previously proposed colliding-jets technique \cite{Kaganovich_15}. This technique utilizes two or more gas jets directed toward a common axis. Such a configuration always generates a density depression on the axis. The radial gas density gradient is responsible for the lensing effect. If the jets are equidistant from the axis, a vortex structure with minimum density at the middle is generated \cite{Kaganovich_15}. This stable configuration is demonstrated and analyzed herein. 

\section{Experiments}

\begin{figure}[htbp]
\centering\includegraphics[width=7.0 cm]{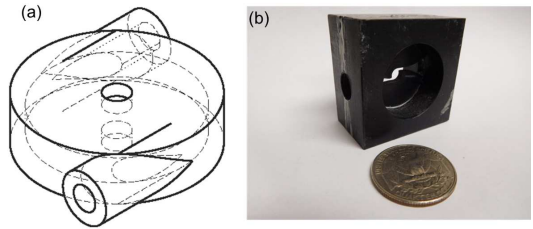}
\caption{(a) Conceptual design of the rotational flow gas lens. Gas enters the lens via inlet tubes in horizontal plane. Gas exits the lens in vertical directions through output holes. (b) 3D printed gas director inside the lens body. Quarter dollar is shown for the scale.}
\label{fig1}
\end{figure}

The conceptual design of the gas lens is depicted in Figure \ref{fig1}a. A continuous flow of compressed gas enters the cylindrical lens body from two opposite sides (inlets) and is forced into rotation inside the central disk of the lens. The rotating gas exits the lens body through outlets along the optical axis that are perpendicular to the inlets plane. The gas exits the lens into an ambient environment that can be atmospheric or a vacuum created by differential pumping. The clear aperture of the gas lens is determined by the diameter of the outlets.

From practical construction considerations, we limited the number of gas inlets to two. A larger number of inlets is possible, but does not provide any significant advantages. For the high intensity down-collimation application the output beam is typically limited to 1-2 mm or even smaller diameters \cite{Sprangle_02}. These beams dictate a typical gas lens size with clear aperture of 1-3 mm. Thickness of the lens was chosen to be 2 mm, close to its diameter to avoid a longer tube, where diverging beam might clip on the walls. The experimental prototype of the lens was partially 3D printed out of ABS plastic and partially machined as shown in Figure \ref{fig1}b. Pressure fitting adapters to standard 1/4 inch gas delivery tubes were glued into the main body. In most experiments we used compressed nitrogen and a manual pressure regulator to adjust the gas flow. To keep the pressure the same on both sides, we split a tube from the pressure regulator into both inlets. 

In the first set of experiments a 532 nm, Gaussian beam profile, continuous 50 mW, slightly divergent laser beam was focused using a 14 cm glass lens. The focal spot was imaged by a microscope objective onto a CCD camera. The gas lens was positioned between the laser and the conventional glass lens 30 cm away from the glass lens. The laser beam diameter at the gas lens location was about 1.5 mm. Without the gas flow in the lens the laser beam focal spot remained unchanged on the CCD. 
By analyzing shift and shape of the laser focal spot profile during the lens operation (gas on) two important requirements to the gas lenses construction were optimized. The first important requirement was the gas lens symmetry. Even small deviations in symmetry of the lens construction or assembly resulted in degradation or instability of the focal spot. 3D printer that was used to make the lens parts (Zortrax M200) could not provide enough resolution on the most critical surfaces. Central holes and surfaces were drilled and polished on a milling machine. The second important aspect of the gas lens operation was design of the gas outlets. We discovered that it is important to allow the gas to move radially, away from the optical axis of the lens. This can be achieved by machining external sides of the outlets into wide (90 degrees or larger) cones. If narrower opening cones are used, a significant amount of gas flows along the optical axis outside the lens, generating turbulence and affecting stability of the focal spot.

When the gas flow was turned on, the focal spot position shifted further downstream from the glass lens. The shift increased as the inlet pressure increased reaching the maximum value of about 2 cm at 40 PSI inlet pressure, when the lens reached the maximum possible flow rate (choking). As a result the focal spot did not move any further for higher pressures. The maximum gas flow depends on the lens geometry, especially the inlet and outlet diameters. Precise position of the focal spot was hard to determine due to a large F-number of the optical system and aberrations produced by the gas lens (see below).   

When nitrogen was replaced by helium at the same pressures, very little or no shift in focal spot position was detected. The refractive index of helium is apparently too low to make any significant change in the laser beam path, and for the rest of this paper, the working gas used in all experiments is nitrogen at 40 PSI inlet pressure. The lens used in experiment is 2 mm thick, with 1.75 mm diameter outlets.

\begin{figure}[htbp]
\centering\includegraphics[width=8.5 cm]{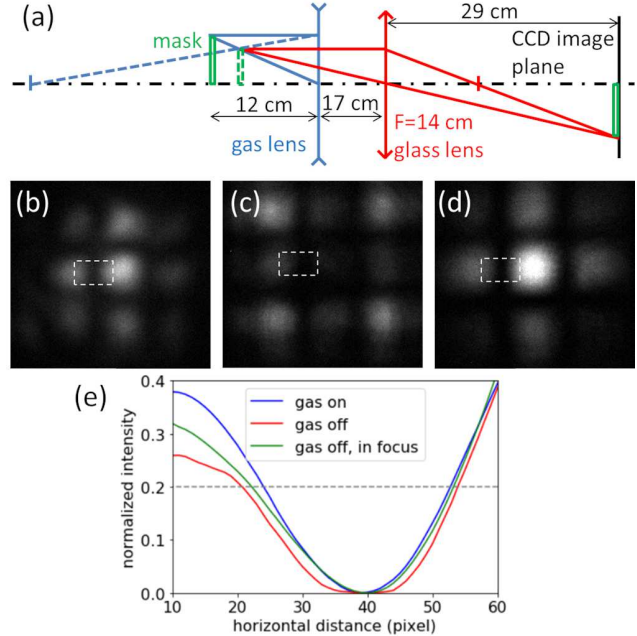}
\caption{(a) Experimental setup and ray tracing for the gas lens focal length measurements. The mask (green rectangle on the left) is imaged onto the CCD image plane through the diverging gas and converging glass lenses. (b) Image of the mask with stationary gas flow (40 PSI inlet pressure) in the lens, (c) image without gas flow for the same optical setup, and (d) sharp image of the mask without gas flow, obtained by adjusting position of CCD camera. This image appears 4\% larger on the CCD than the image in (b), but with similar sharpness. Sharpness of all images was plotted by taking horizontal outlines integrated over areas inside dashed rectangles in (b), (c), and (d) and is shown in (e). Arbitrary chosen dashed gray line in (e) indicates the intensity level where the images sharpness were measured.}
\label{fig2}
\end{figure}

To estimate the focal length of the gas lens we built a simple imaging system shown schematically in Figure \ref{fig2}a. The imaging was done by sending the laser beam through a 270 $\mu$m period mesh mask, the gas lens, and a positive 14 cm glass lens into a microscope objective of a CCD camera. The lenses generated an image of the mask on the camera (Figure \ref{fig2}b). Large F-number of the imaging system was limiting factor of optical resolution defined by 14 cm glass lens and 1.75 mm aperture of the gas lens resulting in 52 $\mu$m radius of the Airy disk - the characteristic size of blurriness in the image. To find the best imaging position of the CCD camera we plotted an outline of a shadow created by the mesh's wire. The sharpest image has the shortest transition between dark and bright pixels of the camera (Figure \ref{fig2}e).

By moving the image plane (CCD camera) back and forth we were able to measure depth of focus of the imaging system. The depth of focus was defined as the distance over which the dark-bright transition width increases by 5\% (twice the smallest detectable level). The width was measured at 0.2 normalized intensity level in Figure \ref{fig2}e. 5\% change in the transition width corresponds to $\pm$5 mm shift of the image plane. Using dimensions and ray traces in Figure \ref{fig2} the focal length of the gas lens is calculated to be -67 $\pm$18 cm, where $\pm$18 cm is the uncertainty due to the depth of focus. When the gas in the lens was turned off, the sharpness of the image was reduced as shown in Figure \ref{fig2}c and outlined in Figure \ref{fig2}e. In order to get the no-gas image back into focus, we moved the imaging plane 2 cm toward the lenses (Figure \ref{fig2}d).

While the gas lens in its present configuration is not designed for imaging, the imaging technique provides a simple way to assess its lensing properties. In addition, the image in Figure \ref{fig2}b indicates aberrations near the edges that are visible when brightness of the images is adjusted to saturation level (not shown). A better way to measure the focal distance and to characterize the aberration would be a wave-front analysis, but it is beyond the scope of this paper and will be subject of future research. 

\section{Simulations}

\begin{figure}[htbp]
\centering\includegraphics[width=8.5 cm]{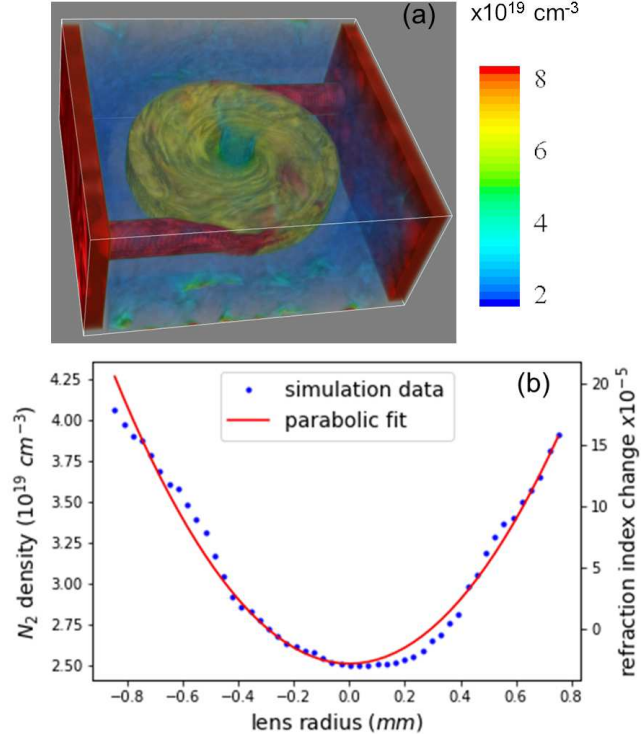}
\caption{ (a) Screen-shot of the nitrogen density in rotational flow inside the gas lens. The diameter of the rotational (yellow) disk of gas is 6 mm, thickness is 1 mm. Thickness of all walls (not shown) is 0.5 mm. Clear aperture of the lens is 1.75 mm. Gas density inside filling chambers (partially shown on each side) is $7.8\times10^{19} cm^{-3}$, corresponds to the inlet pressure of about 40 PSI. (b) Gas density profile across the central horizontal slice inside the clear aperture of the lens. Blue points are simulation values, red line is parabolic fit Eq. \ref{eq1}. The right vertical axis is refractive index change of nitrogen from the standard conditions of 1 atmosphere pressure and 300 K temperature.}
\label{fig3}
\end{figure}

To examine the evolution of the rotational gas flow we used a full 3D, time dependent version of the computer simulation SPARC described in detail elsewhere \cite{Kaganovich_14}. Figure \ref{fig3}a presents a screen-shot of gas density in the gas lens after the flow stabilized. The walls of the lens are not shown; false color map represents high gas densities in red and low gas densities in blue. The simulated lens design and gas flow parameters are set to be similar to experiment shown in Figure \ref{fig2}.
In order to estimate focusing power of the gas lens we have to calculate refractive index change across the clear aperture.  3D simulation shows that all radial gas density profiles inside a cylinder defined by the clear aperture and thickness of the lens look similar to the central slice shown in Figure \ref{fig3}b. The gas density quickly drops down outside of the lens body and does not contribute to the refractive index change. That makes the gas lens similar to a negative gradient-index (GRIN) lens \cite{Smith} that has cylindrical shape with flat bases and a refractive index varying parabolically with radius $r$
\begin{align}
n(r) = n_{min} \left(1+\frac{k_g}{2}r^2 \right)
\label{eq1}
\end{align}
where $n_{min}$ is the refractive index at the center of the lens and $k_g$ is related to the curvature of the parabola. The focal length in the thin-lens limit is
\begin{align}
F = \frac{1}{n_{min}k_gd}
\label{eq2}
\end{align}
where $d$ is thickness of the lens. For a rarefied gas, the quantity $n(r)-1$ is proportional to the gas density $N(r)$. Figure \ref{fig3}b shows that the density profile within the clear aperture of the gas lens is close to parabolic and $k_g$ is found by fitting. The effective thickness of the lens in Figure \ref{fig3} is determined by taking into account the 1 mm thickness of the rotating gas disk and the 0.5 mm of width of top and bottom walls, where the gas flow is restricted from expanding radially. Substituting all parameters into Eq. \ref{eq2} we can estimate the focal length of the lens based on SPARC simulations. The following parameters were found from the parabolic fitting and used in the calculation: $n_{min} = 1.00026$, $k_g = 0.066$ $cm^{-2}$, the lens thickness $d = 0.2$ $cm$. The resulting focal length of -76 cm is within the range found in the experiment. The close agreement between experimental and simulation focal lengths justifies the neglect of boundary layer effects in SPARC runs presented here. The effects of the boundary layer will be addressed in a forthcoming paper.

\section{Summary and outlook}
There are a number of effects that can limit the operation of the lens in practice.  These are thermal blooming constraints at high power and gas breakdown/ionization at high laser intensity. Thermal blooming is caused by heating of the gas due to absorption of laser power.  The result is an increase in the expansion of the laser beam in addition to its normal defocusing property.  The spot radius of the propagating laser beam is affected by the lensing effect of the gas, thermal blooming (provided the intensity is below the breakdown limit) and diffraction. Inside the lens region the last effect is negligible; comparing the contributions to the variation of the spot radius $w$ of the beam one finds that thermal blooming is negligible provided the laser power is below a threshold value given by
\begin{align}
P_{threshold} = \left(\frac{w^2}{F}\right)^2\frac{\pi n^2 N c_p T}{\alpha\left(n^2-1\right)\tau}
\label{eq3}
\end{align}
where $N, c_p, T$ are the density, specific heat at constant pressure and temperature of the gas, respectively, $F$ is the focal length of the gas lens, $\alpha$ is the gas absorption coefficient (in units of inverse length) at this wavelength and $\tau$ is the laser pulse duration \cite{Hafizi_14}.  The expression for the threshold power assumes that the time scales for thermal conduction and for convection/advection of heat are sufficiently long.  On account of the short duration of laser pulses of interest (10’s of ps or less) both of these conditions are readily satisfied.  As an example, for nitrogen gas, based on available information \cite{Hitran}, the absorption coefficient at $\lambda=$532 nm is so small that thermal blooming is negligible for practically all power levels and laser pulse repetition rates of interest here. Additionally, preheating of the gas flowing into the lens will suppress thermal blooming, because the right hand side of Eq. \ref{eq3} is proportional to the gas temperature. Any heat deposited in the lens due to high repetition rates or high average power of the laser is naturally removed by the gas flow and will not affect the lens operation. It should be noted that these considerations assume that self-focusing of the laser beam (Kerr effect) is not significant. That is, the laser power is below the critical power for self-focusing.

The major advantage of the rotational flow gas lens is rapid recovery from any "damage". Ionization by high intensity laser might momentarily disturb optical properties of the lens, but typical intensities for the laser induced ionization in pure gases are higher than in transparent solids. Since one is concerned here with a negative lens that expands the incident beam in the transverse plane one need only consider the possibility of breakdown in the region of incidence. The breakdown intensity depends on the gas and its pressure, the laser wavelength and pulse duration and requires a special study in each specific case \cite{Gray_75}. Generally speaking, however, for pulse lengths in the fs-ps regime breakdown proceeds via multiphoton and/or tunneling ionization.  Multiphoton ionization is prevalent for intensities $\le 10^{12} W/cm^2$, tunneling ionization dominates for intensities $\ge 10^{14} W/cm^2$, and a combination of the two is operative for intensities in between \cite{Sprangle_02}.  For pulses in the ps-ns regime the threshold intensity increases as the pulse length decreases whilst the threshold decreases with increasing pressure \cite{Ireland_74}.  For longer pulse lengths (extending to the $\mu$s) breakdown occurs at relatively low intensities and proceeds via collisional-avalanche ionization and as a function of pressure the lowest breakdown intensity occurs at the Paschen minimum \cite{Raizer_77, Sprangle_11, Penano_12}. The breakdown intensity can be significantly lower in the case of admixtures of gas and/or impurities (eg, aerosols); however, high outflow of gas protects the gas lens from surrounding elements and impurities. For comparison, laser intensities for damages in solids were measured in \cite{Said_95, Couairon_05}.

In addition to a high damage threshold, one of the advantages of the gas lens is the possibility of fast (sub-millisecond) modification of the focal length. In fact, the stationary flow shown in Figure \ref{fig3} simulation was established 200 $\mu$s after the gas was turned on. This is consistent with estimates based on equalization of a pressure change across the lens diameter. This process typically takes a few acoustic times (characteristic length over the speed of sound) and in case of 1 cm lens enclosure is about 100 $\mu$s. Such fast manipulation of the lens parameters requires sub-millisecond control of the gas flow. Recent developments demonstrated sub-millisecond gas injection system \cite{Griener_17} with 300 $\mu$s response time of the gas valve. Incorporating such gas flow control system into the inlet gas line close to the lens body will allow fast manipulation of the gas density inside the lens and as a result fast adjustment of the lens focusing. Experimental demonstration of the focus adjustment is subject of forthcoming experiments. Finally, the gas lens can operate in conditions where conventional optical components are especially sensitive to damage, like marine environment.

Further optimization of the gas lens might require more precise density shaping and different sizes of the lens outlets. That will bring the initial prototype used in our experiments and simulations closer to the ideal lens, free of spherical aberrations \cite{Gordon_18}. Ionization of the gas within the clear aperture will turn the negative lens into a shorter focal length positive lens \cite{Hubbard_02, Hafizi_03}. In case of fully ionized (one electron from each atom) gas, Eq. \ref{eq1} is replaced by
\begin{align}
n_p(r) = n_0\left(1-\frac{k_p}{2}r^2\right),
\label{eq4}
\end{align}
where $n_0=1-\frac{1}{2}\left(\frac{\omega_{p0}}{\omega}\right)^2$, $k_p = \frac{1}{2}\left(\frac{\omega_{p0}}{\omega}\right)^2\frac{n_{min}}{n_{min}-1}\frac{k_g}{n_0}$, $\omega_{p0}=\sqrt{2\frac{4\pi N_{20}e^2}{m}}$ is the on-axis plasma frequency evaluated at twice the minimum nitrogen density $N_{20}$, $\omega$ the optical frequency, and $e$ and $m$ are electron charge and mass. The above calculations were performed with an assumption $\omega_{p0} \ll \omega$. For the parameters considered here, $k_p \approx 24.6 k_g$ and the plasma thin-lens focal length (Eq. \ref{eq2}) is +3 cm. Note that ionization will also increase the damage threshold of the lens by orders of magnitude \cite{Palastro_15}.

This work was supported by the Office of Naval Research.

\end{document}